\documentclass[twocolumn,showpacs,prb,superscriptaddress]{revtex4}
\newcommand{\figurewidth}{\columnwidth}
\usepackage{graphicx}

\newcommand{\av}{_{\mathrm{av}}}

\begin{document}

\title{Defect energy of infinite-component vector spin glasses}

\author{L.~W.~Lee}
\affiliation{Department of Physics,
University of California,
Santa Cruz, California 95064}

\author{A.~P.~Young}
\homepage[Homepage: ]{http://bartok.ucsc.edu/peter}
\email[Email: ]{peter@bartok.ucsc.edu}
\affiliation{Department of Physics,
University of California,
Santa Cruz, California 95064}


\begin{abstract}
We compute numerically the zero temperature defect energy, $\Delta E$,
of the vector
spin glass in the limit of an infinite number of spin components $m$, for a
range of dimensions $ 2 \le d \le 5$.  Fitting to $\Delta 
E \sim L^\theta$, where $L$
is the system size, we obtain: $\theta \simeq -1.54\, (d=2)$, $\theta \simeq
-1.04\, (d=3)$, $\theta \simeq -0.67\, (d=4)$ and $\theta \simeq -0.37\,
(d=5)$. These results show that the lower critical dimension, $d_l,$ (the
dimension where
$\theta$ changes sign) is significantly higher for $m=\infty$
than for finite $m$ (where $2 < d_l < 3$). 
\end{abstract}

\maketitle

\section{Introduction}
\label{sec:introduction}


There has recently\cite{hastings:00,aspelmeier:04,lee:04} been interest in
spin glasses where the number of spin components, $m$, is infinite, because
this limit provides some simplifications compared with Ising ($m=1$), XY
($m=2$), or
Heisenberg ($m=3$) models.  For example, in mean field theory (i.e. for the
infinite range model) there is no ``replica symmetry
breaking''~\cite{almeida:78b} so the ordered state is characterized by a
single order parameter $q$, rather than by an infinite number of order
parameters (encapsulated in a function $q(x)$) which are
needed~\cite{parisi:80} for finite-$m$.  In addition, there are special
numerical techniques\cite{bray:82,morris:86,hastings:00,aspelmeier:04,lee:04}
which can be used to study
finite-range $m=\infty$ spin glasses in which the (finite) sample is solved
\textit{exactly}
without the statistical errors and equilibration problems inherent in
the Monte Carlo methods used for finite-$m$.

There are, however, significant differences between Ising, XY, and Heisenberg
spin glasses, on the one hand, and $m=\infty$ spin glasses on the other.
In the Ising spin glass in three dimensions there is clearly a finite
temperature transition\cite{ballesterosetal:00},
and we have
argued\cite{lee:03} that the same is true for XY and Heisenberg spins, though
the latter is still somewhat controversial, see
e.g.~Refs.~\onlinecite{kawamura:01,kawamura:98,hukushima:00}. 
Hence, for $m=1, 2$ and 3 the lower critical dimension, below which $T_c$
is zero, is \textit{less than} 3 (in fact
$2 < d_l < 3$). However, for $m=\infty$, one 
finds\cite{morris:86,lee:04} $T_c = 0$ in three-dimensions, so $d_l$
must be \textit{greater
than} 3 in this case. In fact, Viana\cite{viana:88} make the surprising
claim that $d_l=8$  for $m=\infty$, by attempting to sum up the
perturbation expansion. Curiously the upper critical dimension (above which the
critical exponents have mean field values) is
\textit{also} predicted\cite{green:82} to
be $d_u = 8$, which is again different from the value for finite $m$ where
$d_u=6$.

In this paper we attempt to determine the lower critical dimension of the
$m=\infty$ spin glass by computing the zero temperature ``defect energy'',
$\Delta E$,
for a range of dimensions, $2 \le d \le 5$. The defect energy is the
characteristic energy change when the boundary conditions are changed from
periodic (say) to antiperiodic\cite{bray:84,mcmillan:84,fisher:86}.
It is expected that 
\begin{equation}
\Delta E \sim L^\theta
\label{scale_dE}
\end{equation}
where
$L$ is the system size and $\theta$ is a ``stiffness exponent''. If $\theta >
0$ then the system is stiff on large lengthscales and so one expects $T_c > 0$
whereas if $\theta < 0$ then it costs very little energy to break up the
ground state configuration at large scales, and so presumably $T_c = 0$. Hence
$d_l$ is the dimension where $\theta = 0$. For the case where $\theta < 0$, so
$T_c = 0$, the correlation length $\xi$ diverges as $T \to 0 $ like
$\xi \sim T^{-\nu}$, and standard scaling arguments\cite{bray:84,mcmillan:84}
then show that $ \nu = - 1 / \theta$.

Our main result is that
$\theta < 0$ for the full range of dimensions ($2 \le d \le 5$) that we are
able to study showing that $d_l$ is significantly greater than 5, i.e.~much
larger than for finite-$m$.

In Sec. II we discuss the models and numerical implementation.
In Sec. III we discuss our results from simulation in two to
five dimensions. We give our conclusions in Sec. IV.

\section{Model and method}
\label{sec:model}
We take the Edwards-Anderson~\cite{edwards:75} Hamiltonian
\begin{equation}
{\cal H} = -\sum_{\langle i, j \rangle} J_{ij} \mathbf{S}_i \cdot \mathbf{S}_j
\, , 
\label{ham}
\end{equation}
where the spins $\mathbf{S}_i$ ($ i = 1, \cdots, N$) are classical vectors
with $m$ components and normalized to length $m^{1/2}$, i.e. $\mathbf{S}_i^2 =
m$. The summation is over nearest-neigbor pairs. The interactions $J_{ij}$
connect nearest neighbors and are independent random variables with a gaussian
distribution with zero mean
and standard deviation unity.

At finite temperature and for $m=\infty$
the spin-spin correlation functions,
\begin{equation}
C_{ij} \equiv {1 \over m} \langle \mathbf{S}_i \cdot \mathbf{S}_j \rangle ,
\end{equation}
are obtained from the following set of
equations\cite{bray:82,morris:86,hastings:00,aspelmeier:04,lee:04},
\begin{eqnarray}
T^{-1} C_{ij} & = & \left(A^{-1}\right)_{ij},  \qquad \mbox{where}
\label{chi} \\
A_{ij} & = & H_i \delta_{ij} - J_{ij} .
\label{A}
\end{eqnarray}
Here the $H_i$ ($i=1,\cdots,N=L^d$) are Lagrange
multipliers enforcing the normalization of the spins:
\begin{equation}
C_{ii} = 1, \qquad (i = 1, 2, \cdots, N) .
\label{Cii}
\end{equation}
To proceed, one solves the $N$ equations, Eq.~(\ref{Cii}), to obtain the
$H_i$, and then determines the $C_{ij}$ from Eqs.~(\ref{chi})
and (\ref{A}). 

At zero temperature,
Eqs.~(\ref{chi}) and Eq.~(\ref{A}) are no longer well defined. However,
since there are no thermal fluctuations, each spin lies parallel to 
its local field:
\begin{equation}
\textbf{S}_i = H_i^{-1}\sum_j J_{ij} \textbf{S}_j .
\label{self_cons_T0}
\end{equation}
Remarkably, it was shown by Hastings\cite{hastings:00} that these local
fields are precisely the zero temperature limit of the $H_i$ in
Eq.~(\ref{A}). Another interesting result found by Hastings is
that the number of
independent spin components $m_0$ used to form the groundstate
satisfies a bound $m_0 < \sqrt{2 N}$. By independent we meant that
we can always define coordinates for the spins
such that the projections of
the spins are only non-zero for $m_0$ directions and no spin
components are found in remaining $m - m_0$ directions. Furthermore,
it is found numerically that
\begin{equation}
[m_0]\av \sim N^\mu , \quad (m_0 < m) ,
\label{mu}
\end{equation}
where $[\cdots]\av$ denotes an average over disorder,
and the values of $\mu$ we find for dimensions between 2 and 5
are given in Table \ref{mu_table}.

We now see that to study
the $m \to \infty$ limit, we simply need that $m$ should
be greater than $m_0$.
Since calculating $m_0$ involves finding the eigenvalues of a
large matrix ($N \times N$) for $N=L^d$, it is computationally
intensive for large $L$ and $d$.
In practice, we can determine $[m_0]\av$ from a smaller range of
sizes and fit to Eq.~(\ref{mu}). This allows us to
extrapolate the value of $[m_0]\av$ to larger sizes. We then
choose
$m$ to be significantly
greater than $[m_0]\av$ for all $L$. Note some tolerance is required because
the precise value of $m_0$ varies from sample to sample.

To find the groundstate, we use a ``spin-quench'' method \cite{morris:86}.
Firstly, the local field on $\mathbf{S}_i$ is computed via
\begin{equation}
H_i  ={1\over m^{1/2}} \left| \sum_j J_{ij} \mathbf{S}_j\right| \, .
\label{Hi}
\end{equation}
Next we set $\mathbf{S}_i$ according to Eq.~(\ref{self_cons_T0}).
This procedure is applied to each spin of the lattice sequentially
until convergence, and then iterated to convergence. Our convergence
criterion is that the magnitude of the change in each spin is less than about
$10^{-7}$.
For finite $m$, this method does not guarantee the groundstate as there
are many solutions to Eq.~(\ref{self_cons_T0}).
However, it works in the $m \to \infty$ limit because there is a
unique solution \cite{bray:81} in that case.

To determine the defect energy $\Delta E$, we
first find the groundstate energy
with periodic boundary conditions,
$E_p$,
for a given set of bonds. Next, we reverse the $L^{d-1}$
bonds which wrap around the system in one direction ($x$ say), i.e.
the bonds that connect sites with $x=1$ to those with $x=L$.
We then obtain the groundstate energy, $E_a$, for
these ``anti-periodic'' boundary conditions. On average, neither periodic or
anti-periodic boundary conditions
is preferred and so we average the \textit{absolute value} of
$E_p - E_a$ over many different configurations of bonds, i.e. the defect
energy is defined to be
\begin{equation}
\Delta E = [\, | E_p - E_a |\, ]\av \, .
\label{dE}
\end{equation}
We expect that
$\Delta E$ scales with $L$ according to Eq.~(\ref{scale_dE}).

\begin{table}
\begin{tabular}{ | c | c | c | c | c | c |}
\hline
\hline
$D$ & $\mu$ & $\theta$ & $\nu = -1/\theta$ & $\nu$ (Ref.~\onlinecite{lee:04})&
$\nu$ (Ref.~\onlinecite{morris:86})\\
\hline
2 & 0.29 & $-1.54\pm0.02$ & $0.65\pm0.01 $ & $0.65\pm0.05$ & $0.65\pm0.02$ \\
3 & 0.33 & $-1.04\pm0.02$ & $0.96\pm0.02 $ & $1.23\pm0.13$ & $1.01\pm0.02$\\ 
4 & 0.35 & $-0.67\pm0.04$ & $1.49\pm0.09 $ &      --       & $1.5 \pm 0.1$ \\
5 & 0.37 & $-0.37\pm0.07$ & $2.70\pm0.51 $ &      --       & -- \\
\hline
\hline
\end{tabular}
\caption{
Estimates of $\mu$ and $\theta=-1/\nu$. The values of $\nu$ are
compared to estimates from finite temperature
simulations\cite{lee:04} and a previous calculation\cite{morris:86}
at $T=0$.
\label{mu_table}
}
\end{table}

\begin{table}
\begin{tabular}{ | r | c | c | c | c |}
\hline
\hline
& \multicolumn{4}{| c |}{Number of samples, $N_\mathrm{samp}$}\\
\hline
$L$ & D = 2  & D = 3  & D = 4  & D = 5 \\
\hline
4   & 1000 & 1000 & 1000 & 1000\\ 
5   & --   & 1000 & 1000 & 2005\\
6   & --   & 1000 & 1000 & 2098\\
7   & --   & 1000 & 1000 & 1943\\
8   & --   & 1000 & 1000 & --  \\
10  & 1000 & --   & 1317 & --  \\
12  & --   & 1000 & 1042 & --  \\
16  & --   & 1000 & --   & --  \\
20  & 1000 & --   & --   & --  \\
24  & --   & 1000 & --   & --  \\
32  & 1000 & --   & --   & --  \\
64  & 878  & --   & --   & --  \\
128 & 547  & --   & --   & --  \\
\hline
\hline
\end{tabular}
\caption{
Number of samples used in the defect energy calculations.
\label{nosample}
}
\end{table}

\begin{center}
\begin{figure}
\includegraphics[width=\figurewidth]{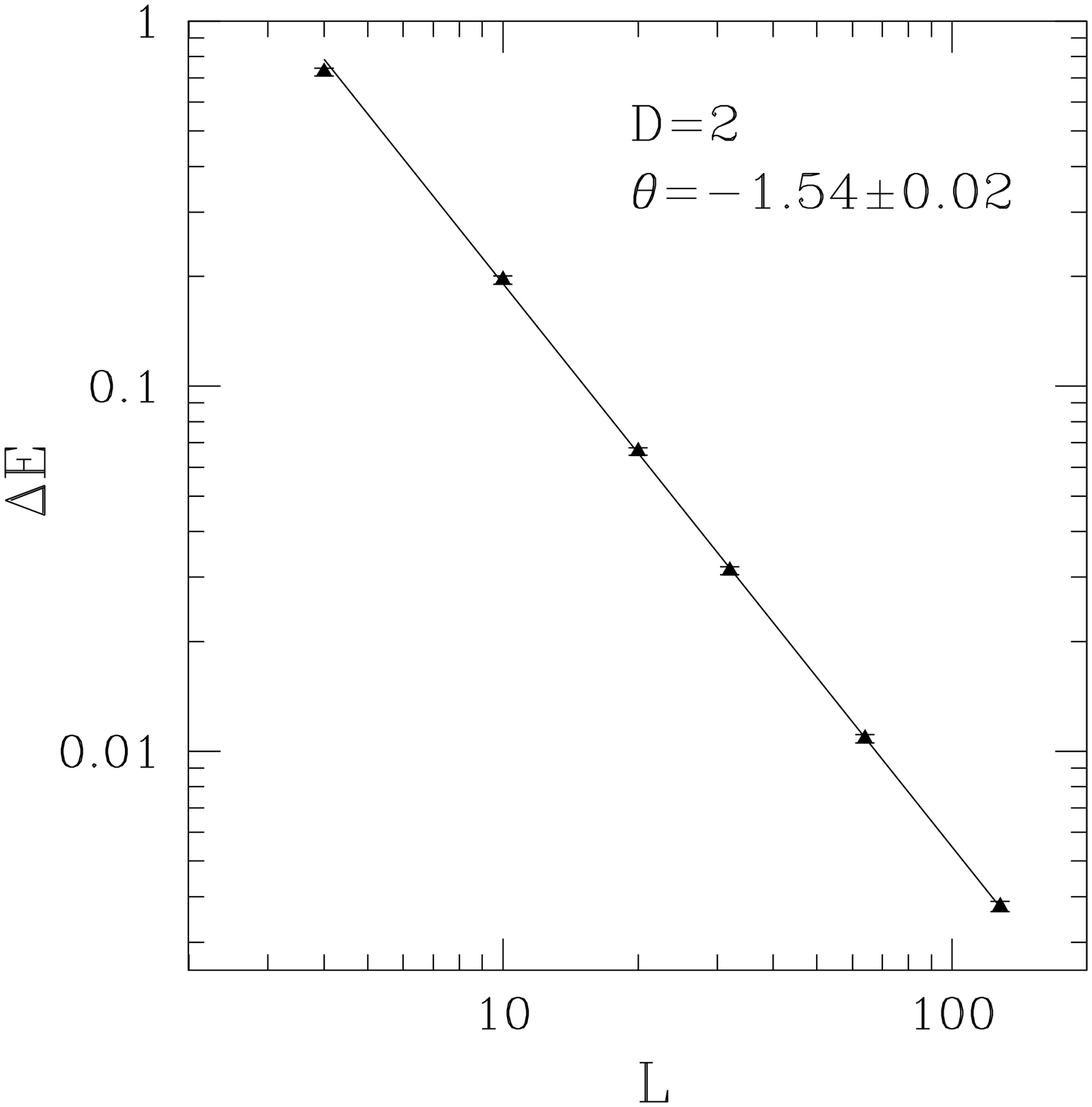}
\caption{
$d=2$: graph of $\Delta E$ against $L$ with $\theta=-1.54$.
}
\label{fig:2D}
\end{figure}
\end{center}

\begin{center}
\begin{figure}
\includegraphics[width=\figurewidth]{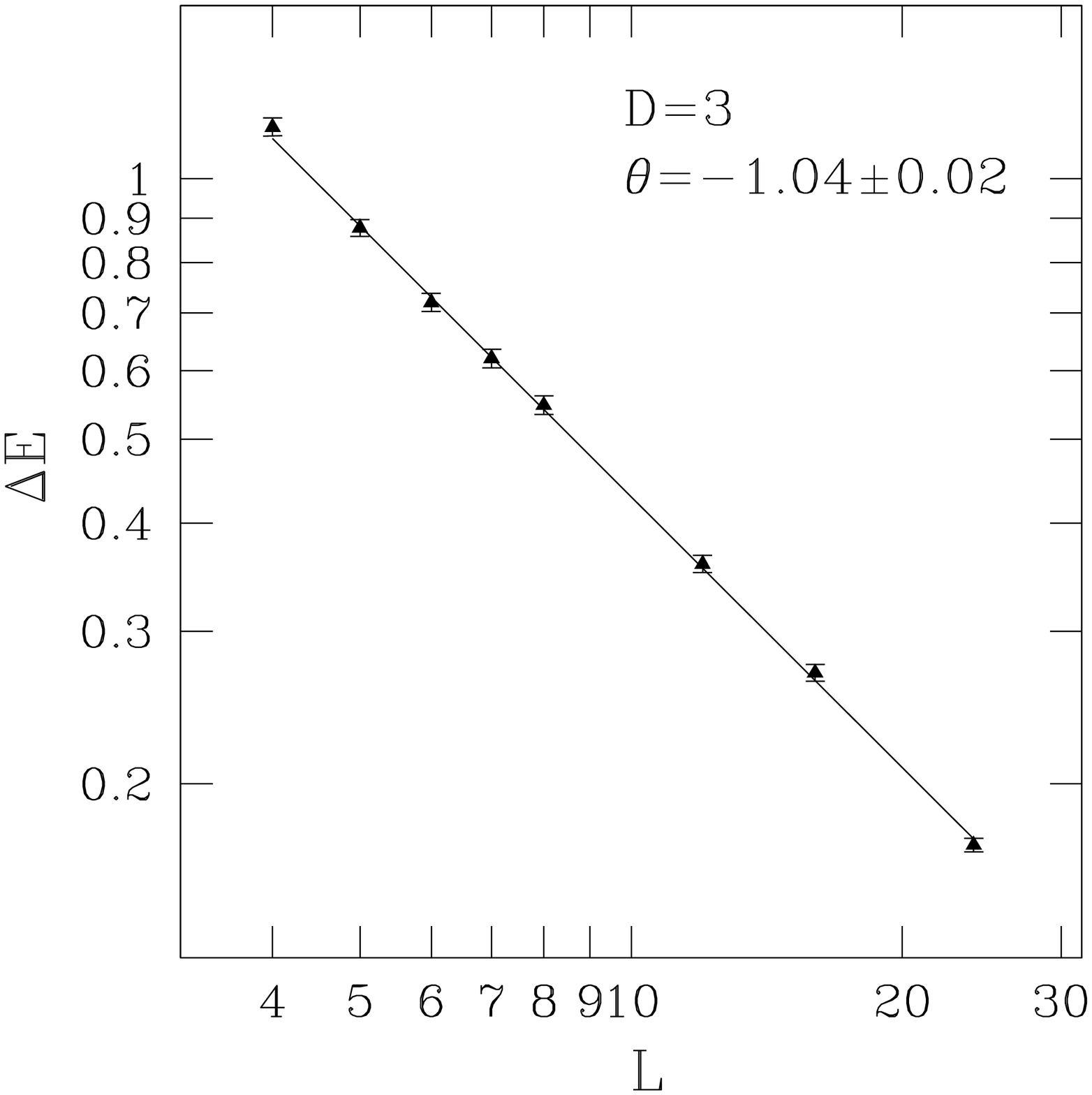}
\caption{
$d=3$: graph of $\Delta E$ against $L$ with $\theta=-1.04$.
}
\label{fig:3D}
\end{figure}
\end{center}

\begin{center}
\begin{figure}
\includegraphics[width=\figurewidth]{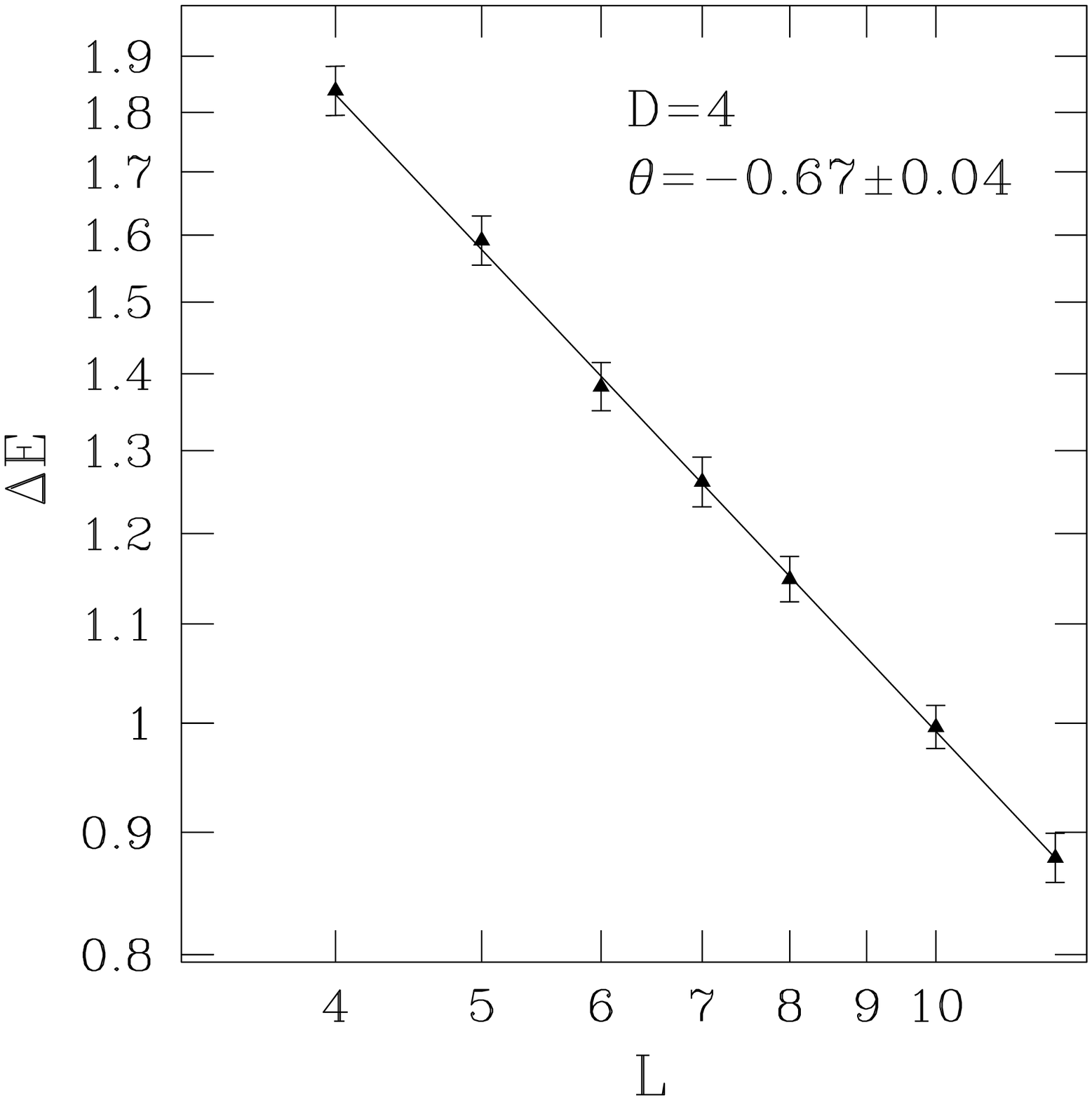}
\caption{
$d=4$: graph of $\Delta E$ against $L$ with $\theta=-0.67$.
}
\label{fig:4D}
\end{figure}
\end{center}

\begin{center}
\begin{figure}
\includegraphics[width=\figurewidth]{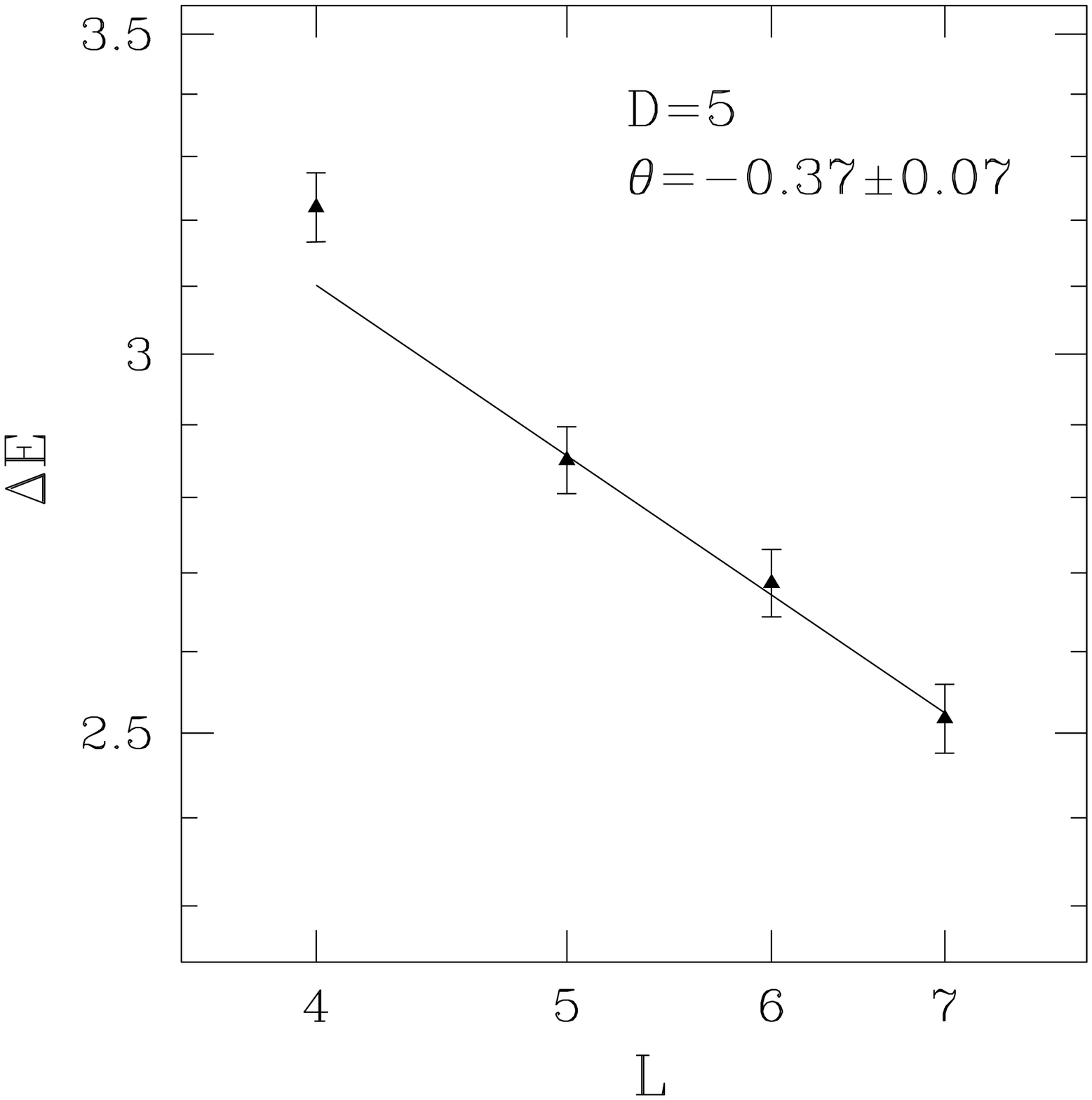}
\caption{
$d=5$: graph of $\Delta E$ against $L$ with $\theta=-0.37$.
}
\label{fig:5D}
\end{figure}
\end{center}

\begin{center}
\begin{figure}
\includegraphics[width=\figurewidth]{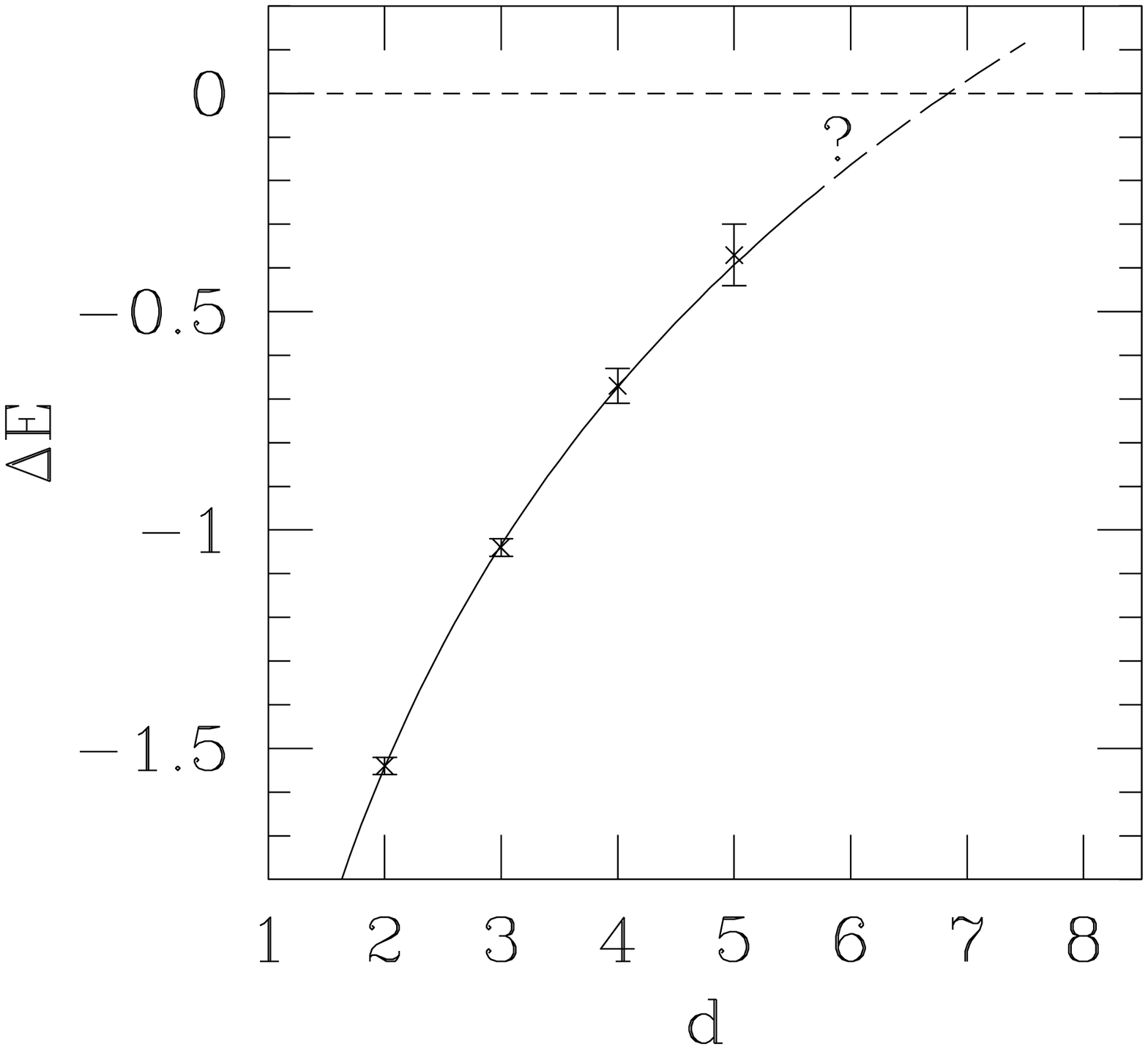}
\caption{
Graph of $\theta$ against $d$. 
The solid line is a smooth curve through the data. Its extrapolation to $d>5$
(curve with long dashes) is,
however, very uncertain.
}
\label{fig:trend}
\end{figure}
\end{center}

\vspace{-3cm}
\section{Results}
\label{sec:results}

We have performed simulations for dimensions $d=$2, 3, 4 and 5. The
number of samples for each size and dimension is presented in
Table~\ref{nosample}. The defect energies for different dimensions
are plotted in Figs~\ref{fig:2D},\ref{fig:3D},\ref{fig:4D}, and
\ref{fig:5D} together with the fit to Eq.~(\ref{scale_dE}).
To reduce finite size effects when doing the fits, for $d=2$ we omit
the two smallest sizes ($L=4$ and 10), and for higher $d$ we
just omit the smallest size ($L=4$).

The exponent $\theta$ thus obtained is shown in Table~\ref{mu_table}
together with that obtained from previous calculations. The results for $d=2,
3$ and 4 agree well with those of Ref.~\onlinecite{morris:86}, though we have
better statistics than in that work and cover a larger range of sizes (128, 24
and 12 as opposed to 12, 7 and 5). We are not aware of any other
results for $d=5$. Comparing with Ref.~\onlinecite{lee:04},
our results for $d=2$ agree very well,
while those for $d=3$ are a little different,
at the level of about $2\sigma$, which may reflect some corrections to scaling.

In Fig.~\ref{fig:trend} we plot our results for $\theta$ as a function of $d$,
together with a smooth curve through the points. It is obviously desirable to
know the dimension, $d_l$, where
$\theta = 0$, but extrapolation of our data to larger $d$ is
very uncertain. However, it is clear that $d_l$ must be significantly greater
than 5, and hence much greater than the its value for finite $m$, which is
between 2 and 3. It is not possible to test precisely the claim of
Viana~\cite{viana:88} that $d_l=8$, because we can't estimate $\theta$ for $d$
close to 8. However, our data does not rule out this possibility.


\section{Conclusions}
\label{sec:conclusions}

We have computed the zero-temperature stiffness exponent $\theta$
for the vector spin
glass in the limit where the spins have an infinite number of components. We
have obtained better statistics for a larger range of sizes and dimensions
that in previous work\cite{morris:86}.  Our results for $\theta$
agree with those of
Ref.~\onlinecite{morris:86} for the dimensions ($d=2$--$4$) considered by
them, while our result for $d=5$ is new. The trend in our data, shown in
Fig.~\ref{fig:trend}, indicates that the lower critical dimension must be
significantly larger than 5, the largest dimension we have been able to study,
and may equal $8$, as predicted by Viana\cite{viana:88}, but it is currently
not technically possible to determine $\theta$ for sufficiently high dimension
to test this claim precisely.

\begin{acknowledgments}
We acknowledge support
from the National Science Foundation under grant DMR No.~DMR 0337049. We also
acknowledge a generous provision of computer time on a G5 cluster
from the Hierarchical
Systems Research Foundation.
\end{acknowledgments}

\bibliography{refs}

\begin{thebibliography}{19}
\expandafter\ifx\csname natexlab\endcsname\relax\def\natexlab#1{#1}\fi
\expandafter\ifx\csname bibnamefont\endcsname\relax
  \def\bibnamefont#1{#1}\fi
\expandafter\ifx\csname bibfnamefont\endcsname\relax
  \def\bibfnamefont#1{#1}\fi
\expandafter\ifx\csname citenamefont\endcsname\relax
  \def\citenamefont#1{#1}\fi
\expandafter\ifx\csname url\endcsname\relax
  \def\url#1{\texttt{#1}}\fi
\expandafter\ifx\csname urlprefix\endcsname\relax\def\urlprefix{URL }\fi
\providecommand{\bibinfo}[2]{#2}
\providecommand{\eprint}[2][]{\url{#2}}

\bibitem[{\citenamefont{Hastings}(2000)}]{hastings:00}
\bibinfo{author}{\bibfnamefont{M.~B.} \bibnamefont{Hastings}},
  \emph{\bibinfo{title}{Ground state and spin glass phase of the large {N}
  infinite range spin glass via supersymmetry}}, \bibinfo{journal}{J. Stat.
  Phys} \textbf{\bibinfo{volume}{99}}, \bibinfo{pages}{171}
  (\bibinfo{year}{2000}).

\bibitem[{\citenamefont{Aspelmeier and Moore}(2004)}]{aspelmeier:04}
\bibinfo{author}{\bibfnamefont{T.}~\bibnamefont{Aspelmeier}} \bibnamefont{and}
  \bibinfo{author}{\bibfnamefont{M.~A.} \bibnamefont{Moore}},
  \emph{\bibinfo{title}{Generalized {B}ose-{E}instein phase transition in
  large-$m$ component spin glasses}}, \bibinfo{journal}{Phys, Rev. Lett.}
  \textbf{\bibinfo{volume}{92}}, \bibinfo{pages}{077201}
  (\bibinfo{year}{2004}).

\bibitem[{\citenamefont{Lee et~al.}(2005)\citenamefont{Lee, Dhar, and
  Young}}]{lee:04}
\bibinfo{author}{\bibfnamefont{L.~W.} \bibnamefont{Lee}},
  \bibinfo{author}{\bibfnamefont{A.}~\bibnamefont{Dhar}}, \bibnamefont{and}
  \bibinfo{author}{\bibfnamefont{A.~P.} \bibnamefont{Young}},
  \emph{\bibinfo{title}{Spin glasses in the limit of an infinite number of spin
  components}}, \bibinfo{journal}{Phys. Rev. E} \textbf{\bibinfo{volume}{71}},
  \bibinfo{pages}{036146} (\bibinfo{year}{2005}), \eprint{cond-mat/0408073}.

\bibitem[{\citenamefont{de~Almeida et~al.}(1978)\citenamefont{de~Almeida,
  Jones, Kosterlitz, and Thouless}}]{almeida:78b}
\bibinfo{author}{\bibfnamefont{J.~R.~L.} \bibnamefont{de~Almeida}},
  \bibinfo{author}{\bibfnamefont{R.~C.} \bibnamefont{Jones}},
  \bibinfo{author}{\bibfnamefont{J.~M.} \bibnamefont{Kosterlitz}},
  \bibnamefont{and} \bibinfo{author}{\bibfnamefont{D.~J.}
  \bibnamefont{Thouless}}, \emph{\bibinfo{title}{The infinite-ranged spin glass
  with m-component spins}}, \bibinfo{journal}{J. Phys. C}
  \textbf{\bibinfo{volume}{11}}, \bibinfo{pages}{L871} (\bibinfo{year}{1978}).

\bibitem[{\citenamefont{Parisi}(1980)}]{parisi:80}
\bibinfo{author}{\bibfnamefont{G.}~\bibnamefont{Parisi}},
  \emph{\bibinfo{title}{The order parameter for spin glasses: a function on the
  interval $0$--$1$}}, \bibinfo{journal}{J. Phys. A.}
  \textbf{\bibinfo{volume}{13}}, \bibinfo{pages}{1101} (\bibinfo{year}{1980}).

\bibitem[{\citenamefont{Bray and Moore}(1982)}]{bray:82}
\bibinfo{author}{\bibfnamefont{A.~J.} \bibnamefont{Bray}} \bibnamefont{and}
  \bibinfo{author}{\bibfnamefont{M.~A.} \bibnamefont{Moore}},
  \emph{\bibinfo{title}{On the eigenvalue spectrum of the susceptibility matrix
  for random systems}}, \bibinfo{journal}{J. Phys. C}
  \textbf{\bibinfo{volume}{16}}, \bibinfo{pages}{L765} (\bibinfo{year}{1982}).

\bibitem[{\citenamefont{Morris et~al.}(1986)\citenamefont{Morris, Colborne,
  Bray, Moore, and Canisius}}]{morris:86}
\bibinfo{author}{\bibfnamefont{B.~M.} \bibnamefont{Morris}},
  \bibinfo{author}{\bibfnamefont{S.~G.} \bibnamefont{Colborne}},
  \bibinfo{author}{\bibfnamefont{A.~J.} \bibnamefont{Bray}},
  \bibinfo{author}{\bibfnamefont{M.~A.} \bibnamefont{Moore}}, \bibnamefont{and}
  \bibinfo{author}{\bibfnamefont{J.}~\bibnamefont{Canisius}},
  \emph{\bibinfo{title}{Zero-temperature critical behaviour of vector spin
  glasses}}, \bibinfo{journal}{J. Phys. C} \textbf{\bibinfo{volume}{19}},
  \bibinfo{pages}{1157} (\bibinfo{year}{1986}).

\bibitem[{\citenamefont{Ballesteros~et al.}(2000)}]{ballesterosetal:00}
\bibinfo{author}{\bibfnamefont{H.~G.} \bibnamefont{Ballesteros~et al.}},
  \emph{\bibinfo{title}{Critical behavior of the three-dimensional {I}sing spin
  glass}}, \bibinfo{journal}{Phys. Rev. B} \textbf{\bibinfo{volume}{62}},
  \bibinfo{pages}{14237} (\bibinfo{year}{2000}),
  \bibinfo{note}{(cond-mat/0006211)}.

\bibitem[{\citenamefont{Lee and Young}(2003)}]{lee:03}
\bibinfo{author}{\bibfnamefont{L.~W.} \bibnamefont{Lee}} \bibnamefont{and}
  \bibinfo{author}{\bibfnamefont{A.~P.} \bibnamefont{Young}},
  \emph{\bibinfo{title}{Single spin- and chiral-glass transition in vector spin
  glasses in three-dimensions}}, \bibinfo{journal}{Phys. Rev. Lett.}
  \textbf{\bibinfo{volume}{90}}, \bibinfo{pages}{227203}
  (\bibinfo{year}{2003}), \eprint{(cond-mat/0302371)}.

\bibitem[{\citenamefont{Kawamura and Li}(2001)}]{kawamura:01}
\bibinfo{author}{\bibfnamefont{H.}~\bibnamefont{Kawamura}} \bibnamefont{and}
  \bibinfo{author}{\bibfnamefont{M.~S.} \bibnamefont{Li}},
  \emph{\bibinfo{title}{Nature of the ordering of the three-dimensional {XY}
  spin glass}}, \bibinfo{journal}{Phys. Rev. Lett.}
  \textbf{\bibinfo{volume}{87}}, \bibinfo{pages}{187204}
  (\bibinfo{year}{2001}), \bibinfo{note}{(cond-mat/0106551)}.

\bibitem[{\citenamefont{Kawamura}(1998)}]{kawamura:98}
\bibinfo{author}{\bibfnamefont{H.}~\bibnamefont{Kawamura}},
  \emph{\bibinfo{title}{Dynamical simulation of of spin-glass and chiral-glass
  orderings in three-dimensional {H}eisenberg spin glasses}},
  \bibinfo{journal}{Phys. Rev. Lett.} \textbf{\bibinfo{volume}{80}},
  \bibinfo{pages}{5421} (\bibinfo{year}{1998}).

\bibitem[{\citenamefont{Hukushima and Kawamura}(2000)}]{hukushima:00}
\bibinfo{author}{\bibfnamefont{K.}~\bibnamefont{Hukushima}} \bibnamefont{and}
  \bibinfo{author}{\bibfnamefont{H.}~\bibnamefont{Kawamura}},
  \emph{\bibinfo{title}{Chiral-glass and replica symmetry breaking of a
  three-dimensional {H}eisenberg spin glass}}, \bibinfo{journal}{Phys. Rev. E}
  \textbf{\bibinfo{volume}{61}}, \bibinfo{pages}{R1008} (\bibinfo{year}{2000}).

\bibitem[{\citenamefont{Viana}(1988)}]{viana:88}
\bibinfo{author}{\bibfnamefont{L.}~\bibnamefont{Viana}},
  \emph{\bibinfo{title}{Infinite-component spin-glass model in the
  low-temperature phase}}, \bibinfo{journal}{J. Phys. A}
  \textbf{\bibinfo{volume}{21}}, \bibinfo{pages}{803} (\bibinfo{year}{1988}).

\bibitem[{\citenamefont{Green et~al.}(1982)\citenamefont{Green, Bray, and
  Moore}}]{green:82}
\bibinfo{author}{\bibfnamefont{J.~E.} \bibnamefont{Green}},
  \bibinfo{author}{\bibfnamefont{A.~J.} \bibnamefont{Bray}}, \bibnamefont{and}
  \bibinfo{author}{\bibfnamefont{M.~A.} \bibnamefont{Moore}},
  \emph{\bibinfo{title}{Critical behavior of an $m$-vector spin glass for
  $m=\infty$}}, \bibinfo{journal}{J. Phys. A} \textbf{\bibinfo{volume}{15}},
  \bibinfo{pages}{2307} (\bibinfo{year}{1982}).

\bibitem[{\citenamefont{Bray and Moore}(1984)}]{bray:84}
\bibinfo{author}{\bibfnamefont{A.~J.} \bibnamefont{Bray}} \bibnamefont{and}
  \bibinfo{author}{\bibfnamefont{M.~A.} \bibnamefont{Moore}},
  \emph{\bibinfo{title}{Lower critical dimension of {I}sing spin glasses: a
  numerical study}}, \bibinfo{journal}{J. Phys. C}
  \textbf{\bibinfo{volume}{17}}, \bibinfo{pages}{L463} (\bibinfo{year}{1984}).

\bibitem[{\citenamefont{McMillan}(1984)}]{mcmillan:84}
\bibinfo{author}{\bibfnamefont{W.~L.} \bibnamefont{McMillan}},
  \emph{\bibinfo{title}{Domain-wall renormalization-group study of the
  three-dimensional random {I}sing model}}, \bibinfo{journal}{Phys. Rev. B}
  \textbf{\bibinfo{volume}{30}}, \bibinfo{pages}{476} (\bibinfo{year}{1984}).

\bibitem[{\citenamefont{Fisher and Huse}(1986)}]{fisher:86}
\bibinfo{author}{\bibfnamefont{D.~S.} \bibnamefont{Fisher}} \bibnamefont{and}
  \bibinfo{author}{\bibfnamefont{D.~A.} \bibnamefont{Huse}},
  \emph{\bibinfo{title}{Ordered phase of short-range {I}sing spin-glasses}},
  \bibinfo{journal}{Phys. Rev. Lett.} \textbf{\bibinfo{volume}{56}},
  \bibinfo{pages}{1601} (\bibinfo{year}{1986}).

\bibitem[{\citenamefont{Edwards and Anderson}(1975)}]{edwards:75}
\bibinfo{author}{\bibfnamefont{S.~F.} \bibnamefont{Edwards}} \bibnamefont{and}
  \bibinfo{author}{\bibfnamefont{P.~W.} \bibnamefont{Anderson}},
  \emph{\bibinfo{title}{Theory of spin glasses}}, \bibinfo{journal}{J. Phys. F}
  \textbf{\bibinfo{volume}{5}}, \bibinfo{pages}{965} (\bibinfo{year}{1975}).

\bibitem[{\citenamefont{Bray and Moore}(1981)}]{bray:81}
\bibinfo{author}{\bibfnamefont{A.~J.} \bibnamefont{Bray}} \bibnamefont{and}
  \bibinfo{author}{\bibfnamefont{M.~A.} \bibnamefont{Moore}},
  \emph{\bibinfo{title}{Metastable states, internal field distributions and
  magnetic excitations in spin glasses}}, \bibinfo{journal}{J. Phys. C}
  \textbf{\bibinfo{volume}{14}}, \bibinfo{pages}{2629} (\bibinfo{year}{1981}).

\end{thebibliography}

\end{document}